\documentclass[aps,prc,preprint,groupedaddress,showpacs]{revtex4-1}
\usepackage{graphicx}
\usepackage[dvips]{color}

\begin{document}

\title{Evaluation of theoretical uncertainties in parity-violating electron scattering from nucleons and nuclei}

\author{O. Moreno}
\author{T. W. Donnelly}
\affiliation{Center for Theoretical Physics, Laboratory for Nuclear Science and Department of Physics, Massachusetts Institute of Technology, Cambridge, MA 02139, USA}

\author{R. Gonz\'alez-Jim\'enez}
\author{J. A. Caballero}
\affiliation{Departamento de F\'{\i}sica At\'{o}mica, Molecular y Nuclear, Universidad de Sevilla, 41080 Sevilla, Spain}

\begin{abstract}
Parity-violating polarized electron scattering from nucleons and nuclei provides an excellent tool to extract valuable information on nuclear and nucleon structure, as well as to determine Standard Model couplings and higher-order radiative corrections. As measurements become more precise, theoretical models should improve accordingly in order to exploit the experimental data fully in extracting meaningful information. At the same time, it is crucial that theoretical evaluations come with realistic estimations of the corresponding theoretical uncertainties to establish that the precision reached in the measurements is not compromised. Here we consider isospin mixing and the charge distribution in nuclei, and strangeness content of the nucleon together with its axial form factor.
\end{abstract}


\maketitle

\section{Introduction}\label{introduction}

Parity-violating (PV) elastic scattering of longitudinally polarized electrons by nucleonic or nuclear targets is an excellent tool to determine the electroweak coupling constants, to test Standard Model features such as the mixing parameters or the role of higher-order radiative corrections, or to explore the neutron distribution in the case of nuclear targets \cite{don89}. We begin by briefly summarizing some of the basic ideas behind such studies.

Parity violation originates in the interference between the vector electromagnetic (EM) current term and the axial weak neutral current (WNC) term of the electron-hadron interaction, these showing opposite behavior under a parity transformation. The degree of violation is usually cast in terms of the PV asymmetry, involving the cross sections of incoming electrons longitudinally polarized parallel ($\sigma^+$) and antiparallel ($\sigma^-$) to their momentum:
\begin{equation}\label{asymmetry_sigmas}
{\cal A}^{PV}=\frac{d\sigma^+ - d\sigma^-}{d\sigma^+ + d\sigma^-}\, .
\end{equation}

By considering a single exchanged boson for each interaction and neglecting the distortion of the electron wave function due to the target Coulomb field (plane-wave Born approximation (PWBA) --- see below), the PV asymmetry is proportional to the ratio of parity-violating to parity-conserving responses:  ${\cal A} = {\cal A}_0 \:W^{PV}/W^{PC}$, with a scale given by ${\cal A}_0 = G_F|Q^2|/(2\sqrt{2}\pi\alpha) \approx 1.8 \cdot 10^{-4} \:|Q^2| $, where $G_F$ and $\alpha$ are the Fermi (weak) and the fine-structure (electromagnetic) coupling constants and the only kinematic dependence is on the four-momentum transfer $|Q^2|$, given here in (GeV/c)$^{2}$. In this report we consider two specific cases of elastic PV electron scattering, {\it viz.}, from protons ($ep$) and from $N=Z$, spin-0, nominally isospin-0 nuclei ($eA$), in the latter case focusing on $^{12}$C as a practical example.

In the case of elastic $ep$ scattering the PV asymmetry can be written as
\begin{eqnarray}
{\cal A}^{PV}_{ep} = {\cal A}_0 \:\frac{a_A\left(\varepsilon G_E^p\widetilde{G}_E^p+\tau G_M^p\widetilde{G}_M^p\right) -a_V\sqrt{1-\varepsilon^2}\sqrt{\tau(1+\tau)}G_M^p G^{ep}_A}{2 (\varepsilon(G_E^p)^2+\tau (G_M^p)^2)}\, , \label{APVep}
\end{eqnarray}
where $a_V$ and $a_A$ represent the vector and axial WNC electron couplings and $G_{E,M}^{p}$ are the electric and magnetic EM form factors of the proton;  we have introduced the kinematical factor $\varepsilon=(1+2\tau(1+\tau)\tan^2(\theta_e/2))^{-1}$ which depends on the scattering angle $\theta_e$, and $\tau\equiv|Q^2|/(4M^2)$ with $M$ the nucleon mass. The WNC form factors can be written as follows (charge symmetry is assumed):
\begin{eqnarray}
 \widetilde{G}_{E,M}^p(Q^2) &=& \xi_V^{p}G_{E,M}^{p} +\xi_V^{n}G_{E,M}^{n} + 
                                \xi_V^{(0)}G_{E,M}^{s} \label{WNCvectorFF}\\
 G^{ep}_A(Q^2)&=&\xi_A^{T=1}G_A^{3} + \xi^{T=0}_A G_A^{8} + \xi_A^{(0)}G_A^{s}\,,
 \label{WNCaxialFF}
\end{eqnarray}
where $G_{E,M,A}^s$ are the strange form factors and $G_A^{3,8}$ the isovector triplet (3) and isoscalar octet (8) contributions to the axial form factor of the proton ($G_A^{ep}$). The $\xi$ coefficients represent the WNC effective coupling constants that are given in terms of the weak mixing angle ($\theta_W$) and radiative corrections (see~\cite{Musolf94} for details).

Parity-violating scattering from the proton is free from nuclear structure ambiguities and can therefore provide detailed information on the WNC nucleon form factors. Our interest in Sec.~\ref{nucleon} concerns the uncertainties associated with the pure EM properties as well as with the axial and vector WNC form factors through the electric and magnetic strangeness contributions and the effective weak couplings. It is clearly important to assess these theoretical uncertainties before drawing definite conclusions concerning higher-order electroweak contributions to the asymmetry.

The $eA$ PV asymmetry is again proportional to the ratio of parity-violating to parity-conserving nuclear responses, with the same scale ${\cal A}_0$ as for the $ep$ case. Under further assumptions, namely elastic scattering from $N=Z$ nuclear targets with pure isospin $T=0$ and angular momentum $J^{\pi}=0^+$ ground states and absence of nucleon strangeness content, the parity-violating and parity-conserving responses are such that the nuclear structure effects cancel out (see \cite{mor14} and references therein) and one gets a reference PV asymmetry:
\begin{equation}\label{referencevalue}
{\cal A}^{PV}_{\text{ref}} \equiv - 2 \:{\cal A}_0 \:a_A \:\sin^2\theta_W \, .
\end{equation}
The extent to which some of the above-mentioned conditions are not actually fulfilled needs to be modeled by theory and extracted from the measured PV asymmetry, so that the remaining effects can be pinned down or the values of the constants accurately determined. This procedure introduces theoretical uncertainties in the analysis due to the spread of results when, for instance, different microscopic nuclear structure models are used, or when different values of the parameters within a model are considered.

To extract new information on electroweak couplings or higher-order interaction effects from $eA$ PV scattering the uncertainty of the PV observables must lie at the few per mil level, typically around 0.3$\%$ \cite{MITworkshop} (both experimental and theoretical uncertainties). On the experimental side, interest has been shown recently in developing relatively low energy, high luminosity polarized electron beams for PV experiments with the hope of reaching the required precision of \cite{MITworkshop} --- proposals such as the MESA accelerator at Mainz \cite{aul11} or an upgraded FEL facility at Jefferson Lab \cite{nei00} go in this direction. On the theoretical side, uncertainties related to the modeling of different nuclear effects are the focus of Sec.~\ref{carbon}.

Having as a goal the basic objectives summarized above, and following the spirit of this JPG Focus Issue,  we discuss how one has attempted to evaluate the theoretical uncertainties in extracting specific information on hadronic form factors, electroweak corrections beyond tree level and nuclear many-body effects such as isospin mixing. We begin in Sec.~\ref{nucleon} with $ep$ PV scattering, followed in Sec.~\ref{carbon} with elastic $eA$ scattering from $^{12}$C and concluding in Sec. \ref{summary} with a brief summary.

\section{Parity-violating electron scattering from the proton}\label{nucleon}

In recent work \cite{Gonzalez-Jimenez13a, Gonzalez-Jimenez14} a systematic analysis of the effects on the PV asymmetry introduced by the various ingredients involved has been presented. Specifically, a variety of prescriptions for the EM form factors, some of them accounting for two-photon exchange effects and other higher-order corrections, was considered. The radiative corrections in the axial form factor and their effects on ${\cal A}^{PV}_{ep}$ were studied in detail, providing results linked to the specific functional dependence with $Q^2$ (value of the axial mass) and to its value at $Q^2=0$. Finally, the sensitivity shown by the PV asymmetry with $s\overline{s}$ (strangeness) content in the nucleon was discussed for both electric and magnetic channels. A comparison of our theoretical predictions for ${\cal A}^{PV}_{ep}$ with available data is also provided.

Let us begin with a few words on one type of uncertainty which is not really a theoretical one, but which affects the extraction of the other ingredients discussed below, {\it viz.} uncertainties introduced by the lack of precise enough knowledge of the {\bf EM form factors} themselves. As seen in Eq. (\ref{APVep}), the PV asymmetry involved an interference of EM with WNC form factors and thus depends on these both via the numerator and denominator in that equation. In principle this source of uncertainty can be controlled by PC electron scattering from the nucleon; however, at present, while knowledge of the EM form factors is very good, nevertheless some uncertainty is inevitable in the PV asymmetry from this source. To summarize, in \cite{Gonzalez-Jimenez13a} it was found that the typical present uncertainty in the asymmetry amounts to  $\sim$2-3$\%$ for $\theta_e=5^{\circ}$, but only $\sim$0.7$\%$ for $\theta_e=170^{\circ}$ for $|Q^2|=1$ (GeV/c)$^2$. However, the experimental situation is not fully settled and with specific data sets the deviation from the average is larger: $\sim$5$\%$ ($\sim$4.3$\%$) for $\theta_e=5^{\circ}$ ($\theta_e=170^{\circ}$) for $|Q^2|=0.8$ (GeV/c)$^2$. These issues are discussed more completely in \cite{Gonzalez-Jimenez13a}, although in the present study we do not pursue this problem, since our main focus is on evaluations of theoretical uncertainties, with which we now continue.

\begin{figure}[htbp]
    \centering
    \includegraphics[width=.35\textwidth,angle=270]{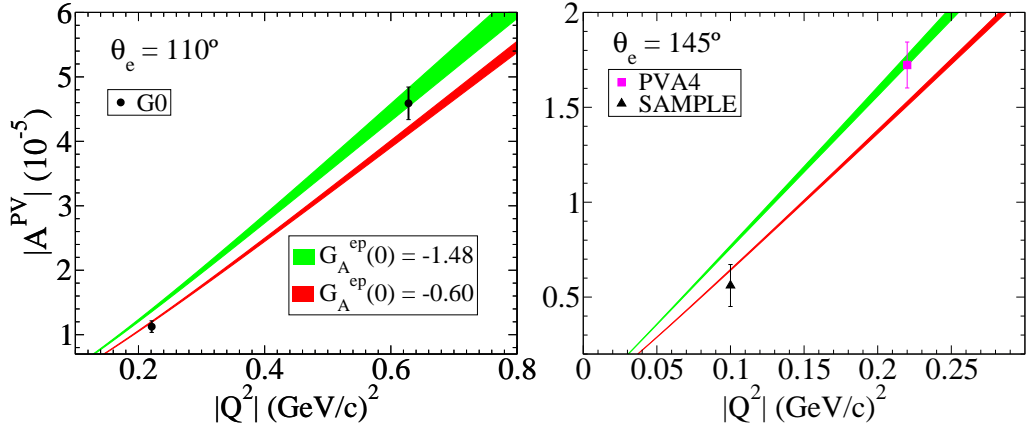}
    \caption{Dependence of the PV asymmetry on the nucleon axial form factor including radiative corrections (see text for details). The upper (lower) extreme line of each band corresponds to $M_A=1.35$ GeV/c$^2$ ($M_A=1.03$ GeV/c$^2$). The magnetic strangeness has been fixed to zero.}\label{Axial2}
\end{figure}

A basic ingredient in the study of PV electron scattering is the WNC {\bf axial form factor} and how its description may modify the results for ${\cal A}^{PV}_{ep}$. It is well known that, contrary to neutrino reactions, radiative corrections play an important role in the description of $G_A^{ep}$ for PV electron scattering. On the other hand, strangeness effects in $G_A^{ep}$ are generally very small.
Thus in the discussion that follows we focus on the effects in ${\cal A}^{PV}_{ep}$ associated with the present knowledge of $G_A^{ep}(Q^2=0)$ and the {\it ``possible''} contributions of radiative corrections. In this work we assume the radiative corrections to be constant. Recently some authors \cite{rad} have studied the possible effects associated with energy-dependence in the $\gamma Z$-box correction. However, at present all calculations have been performed only at very specific kinematics: forward scattering angles and low energy. Although different studies conclude results that differ by some amount, all calculations lead to uncertainties that are much smaller than the ones corresponding to the parameter range
variation considered in this work. Here we follow the analysis presented in~\cite{Gonzalez-Jimenez13a} and show results corresponding to the currently accepted value of the axial form factor at $Q^2=0$, namely, $G_A^{ep}(0)=-1.04\pm 0.44$ (see~\cite{Liu07}). Notice that the value at tree-level, $-g_A =-1.27$, is included within the above range. To analyze how this large variation in $G_A^{ep}(0)$ modifies the PV asymmetry, we compare in Fig.~\ref{Axial2} our predictions for the two extreme values of $G_A^{ep}(0)$ with different data (see \cite{Gonzalez-Jimenez13a, Gonzalez-Jimenez14} for references to the data). The shaded areas represent the uncertainty linked to the particular description of the functional dependence of the axial form factor with the transferred four-momentum $Q^2$.
We assume the usual dipole form with two values of the axial mass: the standard one $M_A=1.032$ GeV/c$^2$ 
(lower limit in the bands) and $M_A=1.35$ GeV/c$^2$ (upper limit). 
The latter is connected with the recent analysis of quasielastic neutrino reactions performed by the MiniBooNE Collaboration~\cite{MiniBooNECC10,MiniBooNENC10}. 
The study of other functional dependences, such as a monopole shape, has been considered in \cite{Gonzalez-Jimenez13a}.
Note that, whereas G0 at lower $|Q^2|$ and the SAMPLE data seem to be consistent with the smaller
(in absolute value) $G_A^{ep}$ (red band), the reverse occurs for G0 at higher $|Q^2|$ and PVA4 data (green band). 
Also, the global uncertainty associated with radiative corrections for $110^{\circ}$ is $\sim$12$\%$ at $|Q^2|=0.75$ (GeV/c)$^2$ and $\sim$9$\%$ at $|Q^2|=0.25$ (GeV/c)$^2$. 
Similar results are found for $145^{\circ}$.
Concerning the role of $M_A$ in the $Q^2$ dependence of the form factor, its global impact depends on the specific value selected for $G_A^{ep}(0)$. 
For $110^{\circ}$ (left panel) the uncertainty shown by the green (red) band is $\sim$3.5$\%$ ($\sim$1.8$\%$) at $|Q^2|=0.25$ (GeV/c)$^2$ and $\sim$6.85$\%$ ($\sim$3.15$\%$) at $|Q^2|=0.75$ (GeV/c)$^2$. Similar comments apply to $145^{\circ}$.

An important objective in studies of PV electron scattering concerns the role of {\bf strange quarks in the electric and magnetic sectors}. At backward angles the electric channel gives a negligible contribution and hence ${\cal A}^{PV}_{ep}$ can give detailed information on the strangeness content in the magnetic form factors. However, notice that in this situation $G_A^{ep}$ (see discussion above) can also play a significant role. At forward angles, where the contribution in the asymmetry from $G_A^{ep}$ is smaller, both the magnetic and electric channels contribute, and accordingly a combined analysis of the two kinematical situations can shed light on the strangeness content in the nucleon. In what follows we present a detailed study comparing our theoretical predictions with all available data up to date, and we provide an estimate of the uncertainty in ${\cal A}^{PV}_{ep}$ due to the strange quark contribution. The $Q^2$-dependence for the strange form factors are taken in their usual form, namely, dipole for $G_M^s(Q^2)$ and dipole times $\tau$ for $G_E^s(Q^2)$. The amount of strangeness content is given through the parameters $\mu_s$ and $\rho_s$ corresponding to the magnetic and electric channels, respectively. 

\begin{figure}[htbp]
    \centering
    \includegraphics[width=.35\textwidth,angle=270]{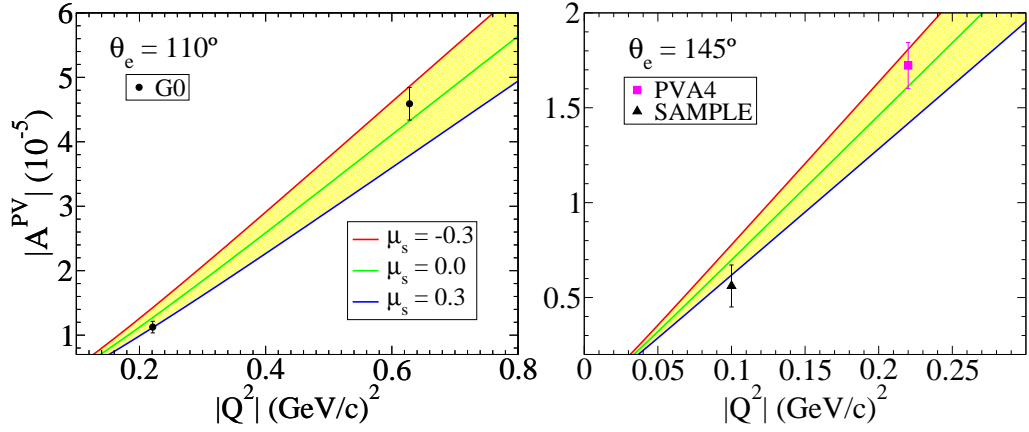}
     \caption{Dependence of the PV asymmetry on the vector magnetic strangeness ($\mu_s$) at backward angles. The value $M_A=1.03$ GeV has been employed.}
        \label{strange-backward}
\end{figure}

\begin{figure}[htbp]
    \centering
    \includegraphics[width=.35\textwidth,angle=270]{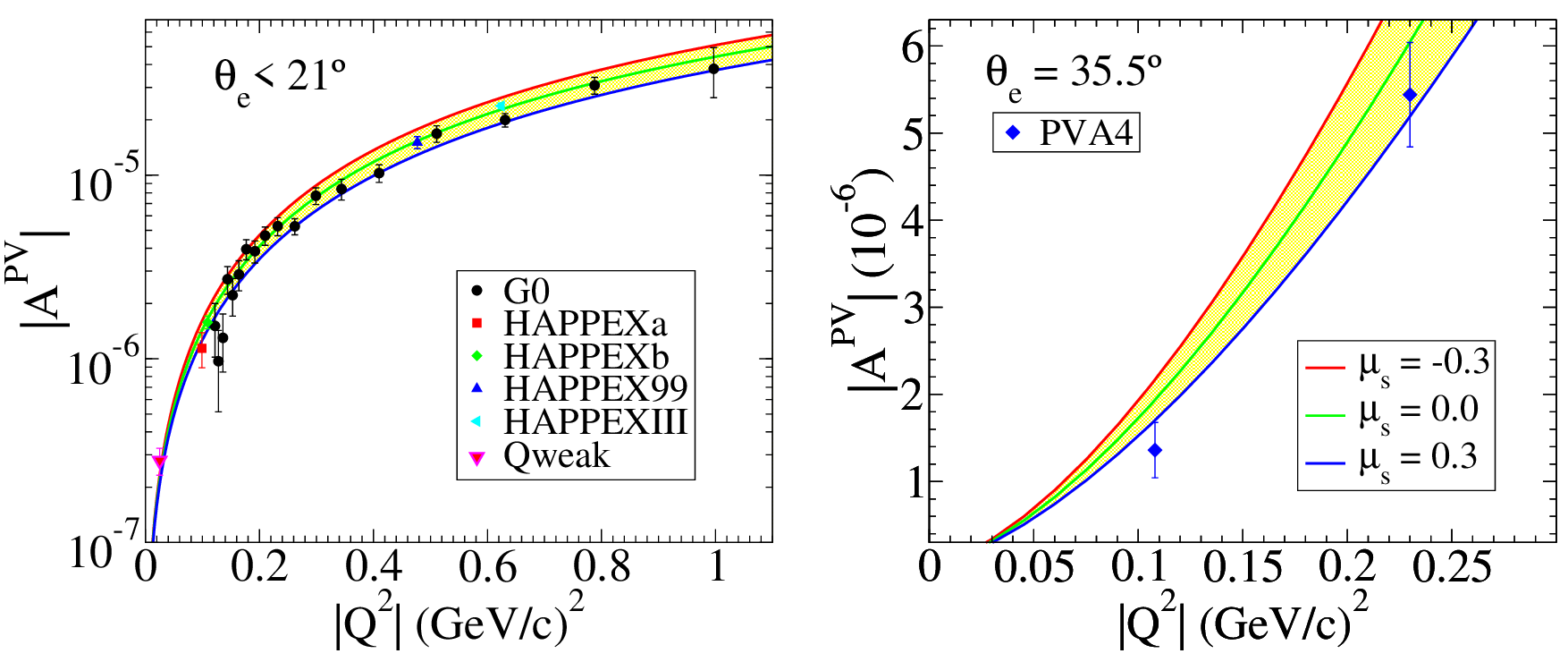}\\
    \includegraphics[width=.35\textwidth,angle=270]{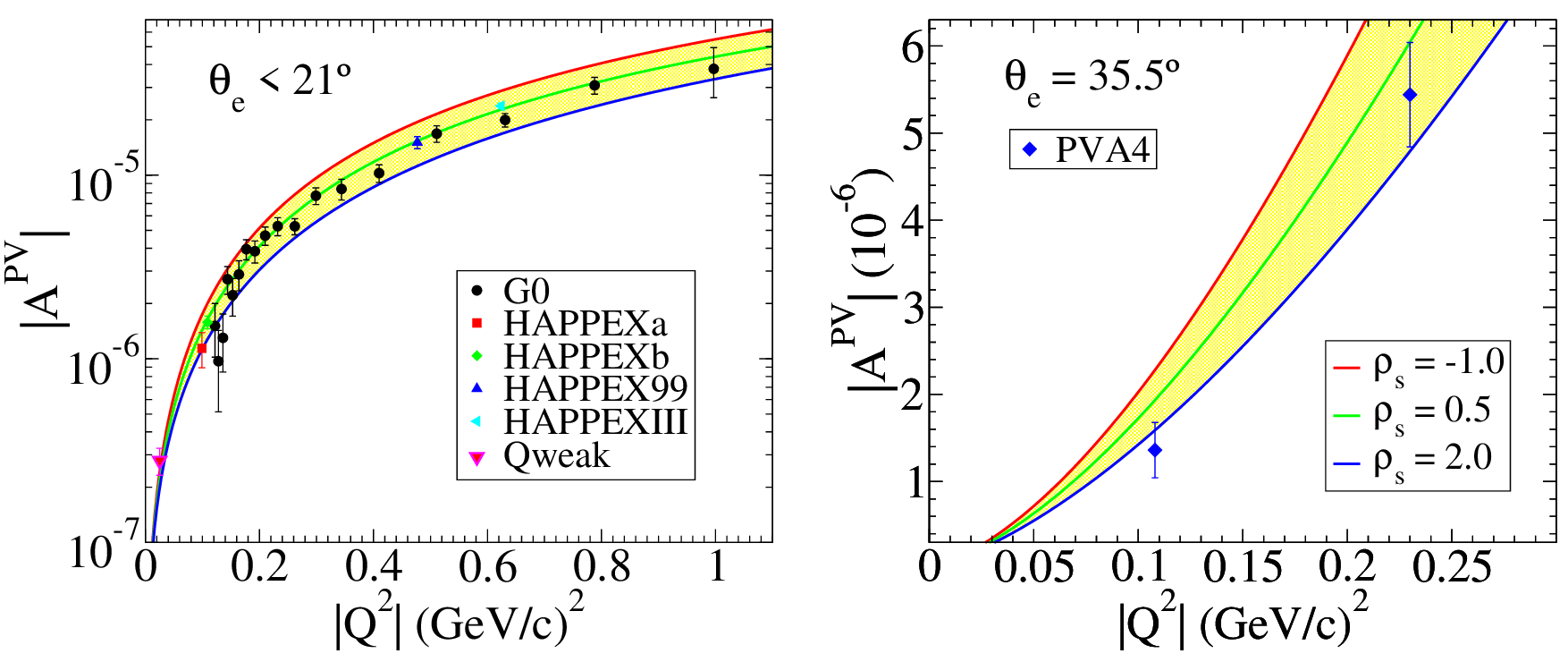}
     \caption{(Top panels) Dependence of the PV asymmetry at forward angles on magnetic strangeness. The electric strangeness has been fixed to $\rho_s=0.5$. (Bottom panels) Dependence of the PV asymmetry at forward angles on electric strangeness. The magnetic strangeness has been fixed to $\mu_s=0.0$.}
        \label{strange-forward}
\end{figure}

Results are shown in Fig.~\ref{strange-backward} (backward angles) and Fig.~\ref{strange-forward} (forward angles). The prescription GKex for the EM form factors has been used (see \cite{Gonzalez-Jimenez13a}) with the dipole-shape description for the standard axial form factor. At backward angles the PV asymmetry depends mainly on the magnetic strangeness content through $\mu_s$. Hence, we show in Fig.~\ref{strange-backward} the band (shaded region) spanned by $\mu_s$-values in the range $[-0.3\, ,\, 0.3]$, which is consistent with previous analyses~\cite{Gonzalez-Jimenez13a}. As observed, the global theoretical uncertainty is large enough to cover the four data analyzed. However, whereas G0 at low $|Q^2|$ and SAMPLE seem to be consistent with positive $\mu_s$, SAMPLE and G0 at higher $|Q^2|$ fit the region of negative $\mu_s$ better. Only the case of magnetic strangeness close to zero seems to be the case where theory and data agree the best. In summary, the uncertainty in the PV asymmetry associated with the variation assumed in $\mu_s$ is of the order of 24--25$\%$. This result applies to the two scattering angles as well as to low and high $|Q^2|$-values.

The case of forward kinematics is analyzed in Fig.~\ref{strange-forward}, where the two panels on the top show the sensitivity of ${\cal A}^{PV}_{ep}$ with respect to the magnetic strange parameter $\mu_s$ with $\rho_s$ (strangeness in the electric channel) fixed. Likewise, the two bottom panels refer to the theoretical uncertainty introduced by the electric strangeness ($\rho_s$) for $\mu_s$ fixed. In both cases the range of variation selected for $\mu_s$ $[-0.3\, ,\, 0.3]$ and $\rho_s$ $[-1.0\, ,\, 2.0]$ is consistent with previous work~\cite{Gonzalez-Jimenez13a}. Moreover, the region spanned by the above selection includes all data and its width (global uncertainty) is slightly larger for the specific $\rho_s$-variation considered. In the case of $\theta_e<21^{\circ}$ (left panels) the width in the band is $\sim$30$\%$ (top panel: magnetic strangeness) and $\sim$53$\%$ (bottom panel: electric strangeness) at $|Q^2|=0.25$ (GeV/c)$^2$. Similar results are found for $|Q^2|=0.8$ (GeV/c)$^2$. For $\theta_e\sim 35.5^{\circ}$ (right panels) the dispersion in the asymmetry is $\sim$26$\%$ (top panel $\rightarrow \mu_s$) and $\sim$39$\%$ (bottom panel $\rightarrow \rho_s$) at $|Q^2|=0.15$ (GeV/c)$^2$. 
In the particular case of the Qweak experiment~\cite{Qweak13}, {\it i.e.,} $\theta_e\approx8^{\circ}$ and $|Q^2|=0.025$ (GeV/c)$^2$, the sensitivity of the PV asymmetry with strangeness results: $\sim$12$\%$ ($\sim$20$\%$) for $\mu_s$ ($\rho_s$).

Finally, let us note that data agree better with theoretical results for positive $\mu_s$ in the case in which $\rho_s$ is fixed to $0.5$ (top panels). Although not shown, such agreement also emerges for negative values of the magnetic parameter, $\mu_s=-0.3$, if the electric strangeness ($\rho_s$-value) is moved to more positive values (region spanned by $\rho_s\in [1.25\, ,\, 2.5]$). This clearly indicates that strangeness in the electric and magnetic sectors are strongly correlated. A similar comment also applies to the axial form factor and its correlation with $G_{E,M}^s$ (see discussion below).

To complete this study we present the results obtained from a {\bf statistical analysis of the full set of PV asymmetry data} for elastic $ep$ scattering including the most recent high precision measurement from Qweak~\cite{Qweak13}. The fit procedure is based on the work presented in~\cite{Gonzalez-Jimenez14} where, in addition to $G_A^{ep}(0)$, $\mu_s$ and $\rho_s$, the proton and neutron WNC effective coupling constants were also included in the analysis. The results obtained for the five free parameters with $1\sigma$-errors are given in~\cite{Gonzalez-Jimenez14}, also showing a strong correlation in most of the cases. As already mentioned, forward angle asymmetry data provide information on the electric and magnetic strangeness while backward angle data are mainly sensitive to the magnetic and axial contributions. Therefore, the strong correlation between the electric ($\rho_s$) and magnetic ($\mu_s$) strangeness parameters comes basically from data taken at forward kinematics (HAPPEX and G0 experiments). In fact, the Jefferson Lab experimentalists often show their results as a specific combination of electric and magnetic strangeness due to this correlation. Likewise, the strong correlation that exists between $\mu_s$ and $G_A^{ep}$ is mainly linked to backward scattering data (SAMPLE, PVA4, and G0 experiments). The statistical analysis presented in \cite{Gonzalez-Jimenez13a} combines experimental data at forward and backward kinematics. This explains the large correlation coefficient (0.711) found also for $\rho_s$ and $G_A^{ep}(0)$ (see \cite{Gonzalez-Jimenez13a,Gonzalez-Jimenez14} for details).

It is worth commenting on the role played by the axial mass connected with the specific functional $Q^2$-dependence of $G_A^{ep}$. This topic was discussed above, showing that the role of $M_A$ is only significant at backward kinematics and for large enough values of $|Q^2|$. Thus, given the current experimental situation with only very few data at backward angles (specifically, the G0 experiment at $|Q^2|\approx 0.63$ (GeV/c)$^2$), the statistical analysis does not seem to be appropriate in order to get clear information on the value of $M_A$. This justifies the use of the standard value $M_A=1.036$ GeV/c$^2$ in the global analysis presented in~\cite{Gonzalez-Jimenez14}.

\begin{figure}[htbp]
    \centering
    \includegraphics[width=.30\textwidth,angle=270]{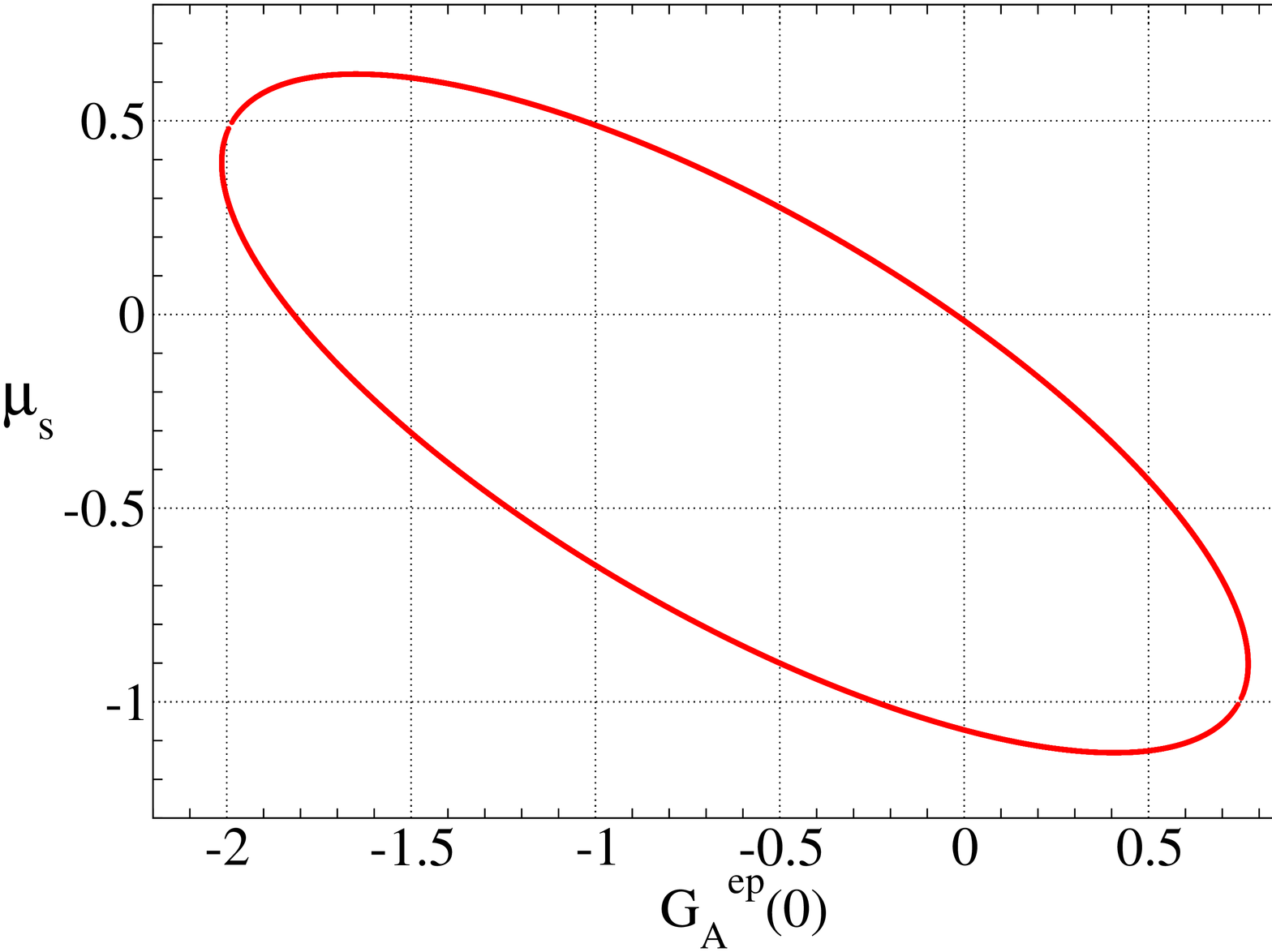}
    \includegraphics[width=.30\textwidth,angle=270]{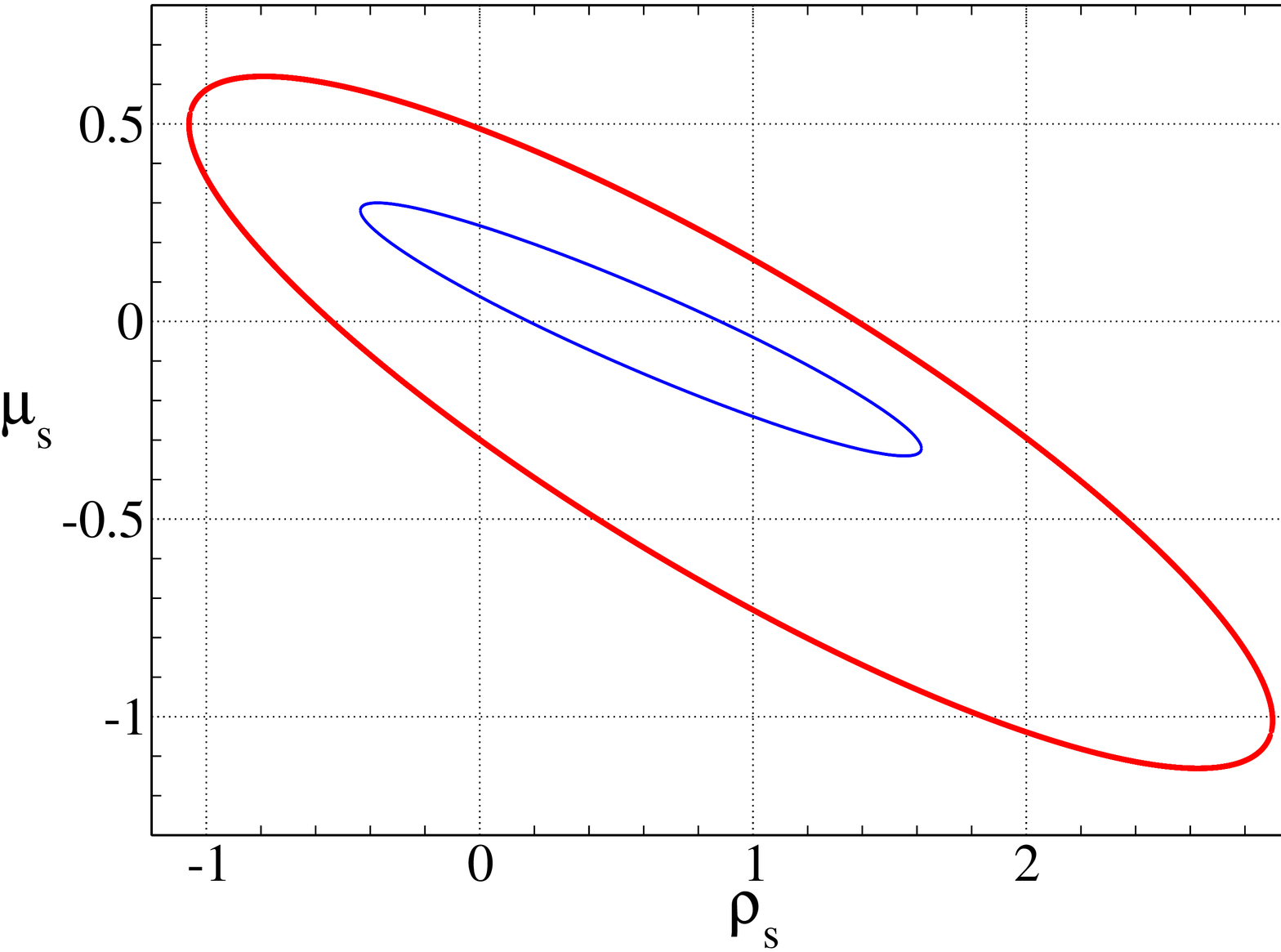}    
     \caption{$95\%$ confidence level contour ellipses in the $[\mu_s,G_A^{ep}(0)]$ (left panel) and in the $[\mu_s,\rho_s]$ (right panel) planes based on the fit procedure presented in~\cite{Gonzalez-Jimenez14} (red curves). The correlation coefficient between $\mu_s$ and $G_A^{ep}(0)$ ($\mu_s$ and $\rho_s$) is -0.749 (-0.870). The blue ellipse in the right hand panel corresponds to the analysis in~\cite{Gonzalez-Jimenez13a}.}
        \label{ellipses}
\end{figure}

In Fig.~\ref{ellipses} we show the $95\%$ confidence level contour ellipses in the $[\mu_s,\rho_s]$ (right) and $[\mu_s,G_A^{ep}(0)]$ (left) planes (red curves). Notice that the values spanned by the ellipses for the three parameters are consistent with the range considered above. Furthermore, the shape of the ellipses indicates the strong correlation existing between the axial form factor and the strangeness content in the nucleon. The blue curve in the right hand panel corresponds to the results obtained in~\cite{Gonzalez-Jimenez13a} where the fit procedure was performed taking only $\rho_s$ and $\mu_s$ as free parameters. This explains the much smaller region spanned by the blue curve compared with the red one (based on five independent parameters).

Finally, although not shown here, the PV asymmetry also shows significant sensitivity to the proton and neutron WNC effective coupling constants. These couplings, that incorporate contributions from higher-order processes ($\gamma Z$-box diagram, $\gamma Z$-mixing diagram, {\it etc.}), were also analyzed in~\cite{Gonzalez-Jimenez14} providing constraints on their values which can be combined with atomic parity-violating measurements. This combined analysis can help in testing the Standard Model predictions (see~\cite{Gonzalez-Jimenez14} and references therein for details).

\section{Parity-violating electron scattering from nuclei}\label{carbon}

We describe next some of the effects that play a role in PV asymmetry of polarized elastic electron scattering from $^{12}$C based on our study in \cite{mor14}. We use experimental conditions similar to those planned for the above-mentioned future facilities, namely 150 MeV longitudinally polarized electrons with a luminosity of $5\cdot 10^{38}$ cm$^{-2}$ s$^{-1}$ scattered by $^{12}$C targets with angles between 25$^{\circ}$ and 45$^{\circ}$, corresponding to momentum transfers between approximately 0.04 and 0.12 GeV/c. We refer to this kinematic interval as the region of interest.

The slightly differing PV asymmetries obtained with different models can be conveniently analyzed through the asymmetry deviation, defined as the relative difference between the theoretical PV asymmetry under study ${\cal A}_X$ (where only the effect $X$ is included) and the reference value ${\cal A}_{\text{ref}}$:
\begin{equation}\label{definition_deviation}
\Gamma_X \equiv {\cal A}_X / {\cal A}_{\text{ref}} - 1 \, .
\end{equation}
The total PV asymmetry contains all of the effects and could be written in terms of all the individual deviations as
${\cal A}_T \approx {\cal A}_{\text{ref}} \:\left(1+\sum_i \Gamma_{X_i}\right)$, where the small interference terms between different effects ($\Gamma_{X_i}\Gamma_{X_j}$) have been neglected. The theoretical uncertainty of a given deviation, $\Delta \Gamma_{X}$, is directly related to the relative theoretical uncertainty of the corresponding asymmetry. For instance, if two different nuclear models $a$ and $b$ describing the nuclear feature $X$ yield a deviation spread $\Delta \Gamma_{X}$, this value also accounts for the relative theoretical uncertainty of the corresponding asymmetry (with respect to ${\cal A}_{\text{ref}}$), since $\Delta \Gamma_X = \Gamma_{X_a} - \Gamma_{X_b} = ({\cal A}_{X_a}-{\cal A}_{X_b})/{\cal A}_{\text{ref}} = \Delta {\cal A}_X/{\cal A}_{\text{ref}}$. 

The reference asymmetry in Eq.~(\ref{referencevalue}) is obtained for plane-wave incoming and outgoing leptons, {\it i.e.} ignoring their interaction with the Coulomb field of the nuclear target. The latter can be accounted for using the {\bf distorted-wave Born approximation (DWBA)} \cite{ruf82}. Although this approximation is accurate enough for many purposes, the actual nuclear charge distribution that should be used in the calculation, which is responsible for the specific Coulomb field of the nuclear target and therefore for the distortion of the lepton waves, does play a role and so needs to be evaluated. The approach here is simply to consider different charge distributions in the nuclear target, all of them compatible with the experimental rms charge radius; the spread of the PV asymmetry deviations obtained with this procedure is then understood as an estimation of the relative theoretical uncertainty in the PV asymmetry due to this effect. A convenient set of charge distributions can be built from a three-parameter Fermi distribution by changing the values of the radius, diffuseness and central-depression parameters (but keeping the corresponding rms charge radii within the experimental range). Obviously, more sophisticated microscopic nuclear models can be used to obtain charge distributions (below, for instance, the use of Skyrme Hartree-Fock calculations for other purposes is discussed). However, it increases considerably the complexity of the calculation without greatly improving the estimation of the uncertainty.

\begin{figure}
\includegraphics[width=0.5\textwidth] {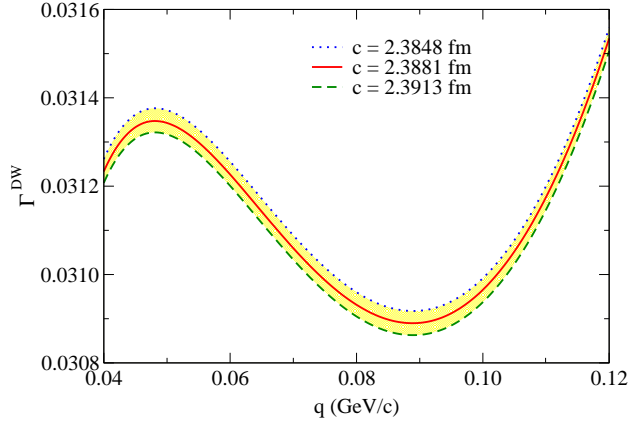}
\caption{PV asymmetry deviation of DWBA results with respect to the reference value as a function of the momentum transfer $q$ for 150 MeV incident electrons on $^{12}$C. The three curves shown correspond to three different values of the radius parameter $c$ of the Fermi charge distribution (indicated in the plot together with the corresponding rms charge radius). \label{dev_dist_e150}}
\end{figure}

We show in Fig.~\ref{dev_dist_e150} the PV asymmetry deviation due to Coulomb distortion effects, $\Gamma_{DW} = {\cal A}_{DW} / {\cal A}_{\text{ref}} - 1$, as a function of the momentum transfer $q$ for 150 MeV incident electrons. As an illustration, three values of the radius parameter $c$ of the Fermi distribution have been used, giving rise to the upper, the central and the lower values of the rms charge radius experimental range; the size of the deviations lies around 3$\%$ and their spread, $\Delta\Gamma$, is lower than 0.01$\%$. Both results are smaller when larger incident electron energies are considered. As a conclusion, Coulomb distortion represents a significant effect that can be removed from the PV asymmetry data keeping the corresponding theoretical uncertainty below the desired goal of about 0.3$\%$. Similar results are obtained when the other parameters of the Fermi distribution are modified. 

\begin{figure}
\centering
\includegraphics[width=0.45\textwidth] {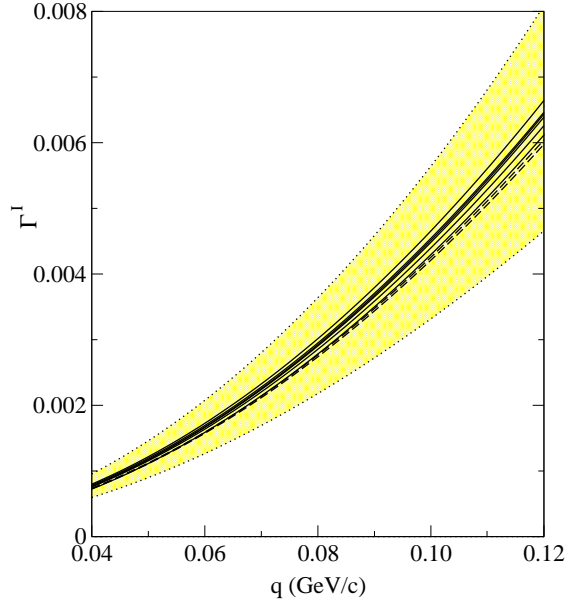}
\caption{PV asymmetry deviation in $^{12}$C due to nuclear isospin mixing with respect to the reference value, as a function of the momentum transfer $q$. Several results are shown for different Skyrme forces used in a Hartree-Fock calculation (thin solid and dashed lines for two groups of similar results, and thin dotted lines for outliers). \label{dev_dist_skyrme}}
\end{figure}

Another condition that was assumed in defining the reference asymmetry of Eq.~(\ref{referencevalue}) is that the $N=Z$ nuclear target has the protons and neutrons following the same dynamics, namely the nuclear target has pure isospin zero. However, at some level one expects to have {\bf isospin mixing}. Certainly, the Coulomb interaction among protons introduces an obvious distinction (maybe not the only one), which translates into a PV asymmetry deviation with respect to the reference value. Details of the proton versus neutron distributions drive this effect, so a rather sophisticated microscopic nuclear model is needed to describe them. We use an axially-deformed Hartree-Fock mean field with Skyrme nucleon-nucleon interactions and pairing correlations via a BCS approximation \cite{mor09}. Skyrme interactions \cite{sky56,vau} are effective, density-dependent nucleon-nucleon interactions that include several contributions with different strength parameters that are usually fitted to reproduce a variety of nuclear properties in different mass regions of isotopes. 
The PV asymmetry deviation to be analyzed in this case compares the asymmetry when isospin mixing is present with the reference value (pure zero isospin), $\Gamma_I = {\cal A}_I / {\cal A}_{\text{ref}}-1$. The theoretical uncertainty in this effect is estimated through the spread of the deviations obtained using a set of representative Skyrme parametrizations in the Hartree-Fock calculation. Fig. \ref{dev_dist_skyrme} shows the isospin deviation as a function of the momentum transfer for a set of different Skyrme parametrizations found in literature (see \cite{dut12} and references therein). Considering the whole set of Skyrme forces without questioning the reliability of any of them, the size of the deviations lies approximately between 0.1 and 0.7$\%$ in the region of interest with a spread between 0.05 and 0.3$\%$, the latter being close to the maximum value acceptable for our purpose.

Other effects have been studied for their impact on the $eA$ PV asymmetry, including uncertainties from {\bf meson-exchange currents (MEC)} and from being unable in projected experimental situations to resolve the ground state and so having to include some number of {\bf excited states} in the total PV asymmetry. Concerning the former, for spin-0 nuclei the MEC effects arise only in the EM and WNC monopole matrix elements and thus are expected to be small at the low momentum transfers of interest. For the latter, the weighting of the inelastic excitations versus the elastic is suppressed essentially by a factor of $1/Z^2$, and accordingly having a few excited states included does not incur a significant uncertainty. The case of $^{12}$C is a good one: there the first excited state ($2^+$) is known to have a suppressed transverse E2 contribution and thus effectively yields the same asymmetry as the elastic scattering; the second excited state is  $0^+$ and thus is C0 and has the same asymmetry as the elastic scattering. Hence, future experiments with resolutions of about 5—10 MeV should not suffer from nuclear structure uncertainties at the few per mil level. These issues are both discussed in more detail in \cite{mor14}.

To conclude this section we study the uncertainty stemming from the content of {\bf strange flavored quarks} within the nucleons (see above), which adds an extra isoscalar term to the weak neutral current not considered in the reference asymmetry of Eq.~(\ref{referencevalue}). As discussed in Sec.~\ref{nucleon}, the nucleon strangeness content is described by electric and magnetic strangeness form factors, each depending in turn on the strangeness content parameters $\rho_s$, $\mu_s$ introduced in Sec.~\ref{nucleon}. We study this effect through the deviation relating the asymmetry that includes strangeness with the reference value (no strangeness), $\Gamma_s = {\cal A}_s / {\cal A}_{\text{ref}} - 1$, and estimate the uncertainty through the spread in deviations using the current experimental ranges given in Sec.~\ref{nucleon} for both the electric and the magnetic strangeness content parameters (the former being the most relevant for elastic scattering from $^{12}$C).

\begin{figure}
\includegraphics[width=0.45\textwidth]{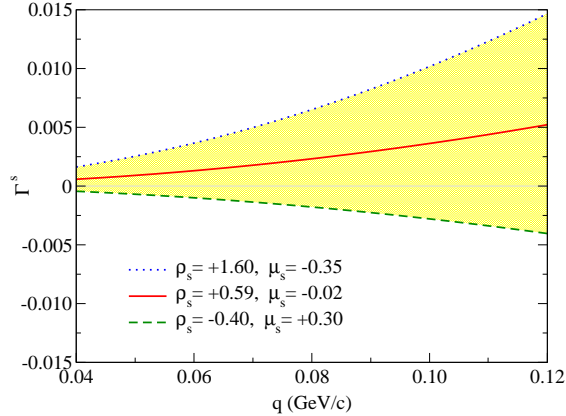}
\caption{Deviation of the PV asymmetry in $^{12}$C due to the strangeness content in the nucleon with respect to the reference value, as a function of the momentum transfer $q$. The three curves correspond to the upper limit, the lower limit and the central value of the experimental range of the electric $\rho_s$ and magnetic $\mu_s$ nucleon strangeness content parameters. \label{dev_strange}}
\end{figure}

In Fig.~\ref{dev_strange} we show the range of deviations --- the overall effect goes from 0.2 to 1$\%$ in the kinematic region of interest and the spread of the results lies between 0.5 and 1.5$\%$, clearly exceeding the maximum uncertainty desired. For the results shown in the figure we have assumed that the uncertainty in
the strangeness content arises from the 2-parameter analysis of PV $ep$ scattering, namely, from the inner ellipse in the right hand panel of Fig.~\ref{ellipses}. Were we to use the 5-parameter analysis, the outer ellipse, then accordingly the spread seen in Fig.~\ref{dev_strange} would be much larger. Therefore, additional asymmetry measurements, ideally for a set of different momentum transfers, seem to be required in order to pin down the strangeness deviation uncertainty. The HAPPEX-He experiment \cite{ach07}, that has measured the PV asymmetry in $^4$He with a 4$\%$ precision at $q=$ 0.277 GeV/c, goes in this direction and suggests that the spread shown in Fig.~\ref{dev_strange} may be correct, but further improvement is still needed.

\section{Summary}\label{summary}

From the discussions above one can identify two types of theoretical uncertainties. On the one hand, when different models exist that describe a given physical feature and there are no compelling reasons to prefer one over the other, an estimation of the theoretical uncertainty could be obtained from the spread of results given by some (or all) of them. One should notice that this is a rather crude approach, since not every model is equally reliable for the description of a given feature, and the results should be weighted accordingly in the estimation of uncertainties (and of the best value). Reliability of a model, without experimental information, is something very difficult to assess, since intrinsically more sophisticated approaches (for instance, a relativistic model versus a non-relativistic one) could lack phenomenological inputs that a less sophisticated theory could include. On the other hand, within a given model or framework, the values of the parameters of the theory could vary within a given range compatible with experimental results. An exploration of the full parameter space compatible with experimental data can result in a reasonable estimation of the theoretical uncertainty of a quantity extracted within the model.

Obviously, experimental uncertainties lie behind both types of theoretical
uncertainties; essentially, if different models or ranges of parameter values
within a model exist, it may be partly because current experimental data are not accurate
enough to pin them down. In addition, experimental and theoretical uncertainties are
sometimes entangled, as is the case of the strangeness content of the nucleon as it
is used in this work: its `experimental uncertainty' actually comes partially from
the different theoretical models that can be used in its extraction from experimental
data (for example, from other studies of the form factors), and this `experimental uncertainty' is subsequently
incorporated as a parameter range in the model of the strangeness form factor.

Let us end by contrasting the two types of reactions discussed in the present paper, namely, $ep$ and $eA$ PV electron scattering. At first one might expect the former to be less subject to modeling uncertainties than the latter, since, clearly, in general it is hard to deal with nuclear structure issues when very high precision is demanded. However, the case of interest here is a very special one: the main focus for $eA$ scattering has been placed on elastic scattering from $N=Z$, nominally isospin-0 nuclei. There, having only a single monopole form factor, the reference PV asymmetry is {\em independent} of any nuclear structure uncertainties, and it is only when one proceeds beyond the na\"ive starting point that these enter. Specifically, when one takes into account that the nuclear ground states of such nuclei are not at some level eigenstates of isospin and therefore that small isovector contributions must be included, then some relatively weak dependence on structure issues arises. Also, when possible strangeness contributions in the nucleons in the nucleus are considered, there are new dependences to take into account. In contrast, the nucleon with its spin-1/2 and isospin-1/2 nature, while more fundamental, nevertheless has a more complicated dependence on the underlying hadronic structure via the various form factors that enter (nine, in fact), none of which is known as well as one might like. Indeed, we have not even considered that it also is likely not an eigenstate of isospin at some level and, furthermore, that both the nucleon and nuclei are not perfect parity eigenstates, which introduces other structure issues --- presumably these are small, and accordingly we have neglected them in this brief analysis. Finally, the `doability' (figure-of-merit) for $eA$ PV forward-angle elastic scattering is considerably larger than it is for $ep$ scattering (see \cite{Musolf94}, Fig.~3.10, and \cite{mor14}), which means that parts per mil determinations of the asymmetry in the former case can be contemplated.

\begin{acknowledgments}
This research was supported by a Marie Curie International Outgoing Fellowship within the 7th European Community Framework Programme, by MINECO (Spain) under Research Grants Nos. FIS2011-23565 (O. Moreno) and FIS2011-28738-C02-01 (R. Gonz\'alez-Jim\'enez and J.A. Caballero), by the Junta de Andaluc\'{\i}a
(FQM-160) and the Spanish Consolider-Ingenio 2000 programmed CPAN. R.G.J. also acknowledges financial support from VPPI-US (Universidad de Sevilla). Also supported in part (T. W. Donnelly) by the US Department of Energy under cooperative agreement DE-FC02-94ER40818.
\end{acknowledgments}


\begin{thebibliography}{19}

\bibitem{don89}
T. W. Donnelly, J. Dubach and I. Sick, \emph{Nucl. Phys. A} \textbf{503}, 1989, pp. 589.

\bibitem{Musolf94}
M. J. Musolf \emph{et al.}, \emph{Phys. Rep.} \textbf{239}, 1994, pp. 1.

\bibitem{mor14}
O. Moreno and T. W. Donnelly, \emph{Phys. Rev. C} \textbf{828}, 2014, pp. 015501.

\bibitem{MITworkshop}
T. W. Donnelly and O. Moreno, \emph{AIP Conf. Proc.} \textbf{1563}, 2013, pp. 82.

\bibitem{aul11}
K. Aulenbacher, \emph{Hyperfine Interact.} \textbf{200}, 2011, pp. 3.

\bibitem{nei00}
G. R. Neil \emph{et al.}, \emph{Phys. Rev. Lett.} \textbf{84}, 2000, pp- 662; D. Douglas \emph{et al.}, Proceedings of the 2001 Particle Accelerator Conference, Chicago, 2001, pp. 249.

\bibitem{Gonzalez-Jimenez13a}
R. Gonz{\'a}lez-Jim{\'e}nez, J. A. Caballero and T. W. Donnelly, \emph{Phys. Rep.} \textbf{524}, 2013, pp. 1.

\bibitem{Gonzalez-Jimenez14}
R. Gonz{\'a}lez-Jim{\'e}nez, J. A. Caballero and T. W. Donnelly, \emph{Phys. Rev. D} \textbf{90}, 2014, pp. 033002.

\bibitem{rad}
M. Gorchtein and C.J. Horowitz, \emph{Phys. Rev. Lett.} \textbf{102}, 2009, pp. 091806; A. Sibirtsev, P.G. Blunden, W. Melnitchouk, and A.W. Thomas, \emph{Phys. Rev. D} \textbf{82}, 2010, pp. 013011; M. Gorchtein, C.J. Horowitz, and M.J. Ramsey-Musolf, \emph{Phys. Rev. C} \textbf{84}, 2011, pp. 015502; N.L. Hall, P.G. Blunden, W. Melnitchouk, A.W. Thomas, and R.D. Young, \emph{Phys. Rev. D} \textbf{88}, 2013, pp. 013011.

\bibitem{Liu07}
J. Liu, R. D. McKeown and M. J. Ramsey-Musolf, \emph{Phys. Rev. C} \textbf{76}, 2007, pp. 025202.

\bibitem{MiniBooNECC10}
A. A. Aguilar-Arevalo \emph{et al.}, \emph{Phys. Rev. D} \textbf{81}, 2010, pp. 092005.

\bibitem{MiniBooNENC10}
A. A. Aguilar-Arevalo \emph{et al.}, \emph{Phys. Rev. D} \textbf{82}, 2010, pp. 092005.

\bibitem{Qweak13}
D. Androic \emph{et al.}, \emph{Phys. Rev. Lett.} \textbf{111}, 2013, pp. 141803.

\bibitem{ruf82}
G. Rufa, \emph{Nucl. Phys. A} \textbf{384}, 1982, pp. 273.

\bibitem{mor09}
O. Moreno, P. Sarriguren, E. Moya de Guerra, J. M. Udias, T. W. Donnelly, and I. Sick, \emph{Nucl. Phys. A} \textbf{828}, 2009, pp. 306.

\bibitem{sky56}
T. H. R. Skyrme, \emph{Phil. Mag.} \textbf{1}, 1956, pp. 1043-1054; \emph{Nucl. Phys.} \textbf{9}, 1959, pp. 615.

\bibitem{vau}
D. Vautherin and D. M. Brink, \emph{Phys. Rev. C} \textbf{5}, 1972, pp. 626; D. Vautherin, \emph{Phys. Rev. C} \textbf{7}, 1973, pp. 296.

\bibitem{dut12}
M. Dutra, O. Lourenco, J. S. Sa Martins, A. Delfino, J. R. Stone, and P. D. Stevenson, \emph{Phys. Rev. C} \textbf{85}, 2012, 035201 pp. 1.

\bibitem{ach07}
A. Acha \emph{et al.}, \emph{Phys. Rev. Lett.} \textbf{98}, 2007, 032301 pp. 1.



\end{thebibliography}
\end{document}